\documentclass[11pt]{article}
\usepackage{epsfig}
\usepackage{amsmath}
\usepackage{amsfonts}
\usepackage{amssymb}

\numberwithin{equation}{section}

\title{Twisted $GL_n$  Loop Group Orbit and Solutions of the WDVV Equations}
\author{Johan van de Leur\\
\\
Mathematical Institute,\\
University of Utrecht,\\
P.O. Box 80010, 3508 TA Utrecht,\\
The Netherlands\\
e-mail: vdleur@math.uu.nl
}

\newtheorem{lemma}{Lemma}[section]
\newtheorem{proposition}{Proposition}[section]
\newtheorem{theorem}{Theorem}[section]

\begin{document}

\maketitle

\begin{abstract}
We show that all (n-component) KP tau-functions, which are related to the
twisted loop group of $GL_n$, give solutions of the Darboux-Egoroff system of
PDE's. Using the Geometry of the Grassmannian we construct from the
corresponding wave function the deformed flat coordinates of the Egoroff metric
and from this the corresponding solution of 
the Witten--Dijkgraaf--E. Verlinde--H. Verlinde equations
\end{abstract}

\section{Introduction}
In the early 90's B. Dubrovin \cite{Du1} noticed that the local classification
of massive topological field theories can be solved by classifying certain flat
diagonal metrics
\begin{equation}
\label{f01}
ds^2=\sum_{i=1}^n h^2_i(u) (du_i)^2,\qquad u=(u_1,\dots,u_n),
\end{equation}
with
\begin{equation}
\label{f01a}
\partial_j h_i^2(u)=\partial_ih_j^2(u),\ \text{and
}\sum_{k=1}^n\partial_kh_i(u)=0,
\qquad \partial_i=\frac{\partial}{\partial u_i}.
\end{equation}
For the flat coordinates $t^i$, $1\le i\le n$, of the metric (\ref{f01}), the
functions
\begin{equation}
\label{f04}
c_{k\ell}^m(t) = \sum_{i=1}^n \frac{\partial t^m}{\partial u_i} \frac{\partial
u_i}{\partial t^k}
\frac{\partial u_i}{\partial t^\ell},
\end{equation}
satisfy
\begin{equation}
\label{f05}
\sum_{k=1}^n c_{ij}^k(t) c_{km}^\ell(t) = \sum_{k=1}^nc_{jm}^k(t)
c_{ik}^\ell(t) .
\end{equation}
If one writes down these equations for the function $F(t)$ for which
\begin{equation}\label{f01b}
\frac{\partial^3 F(t)}{\partial t^k \partial t^\ell \partial t^m}=c_{k\ell
m}(t)=\sum_{i=1}^n\eta_{mi} c_{k\ell}^i(t),\quad
\mbox{where}\quad \eta_{pq}=\sum_{i=1}^n h_i^2(u) \frac{\partial u_i}{\partial
t^p}
\frac{\partial u_i}{\partial t^q},
\end{equation}
with the constraint
\[
\frac{\partial^3 F(t)}{\partial t^1 \partial t^\ell \partial t^m}=\eta_{\ell
m},
\]
one obtains the well-known the Witten-Dijkgraaf-E. Verlinde-H. Verlinde
(WDVV)-equations \cite{W1}, \cite{DVV}.

Vanishing of the curvature of these metrics (\ref{f01}) can be written  in the
form of  a system of partial differential equations in the cannonical
coordinates $u_i$ for the rotation coefficients
\begin{equation}
\label{f02}
\gamma_{ij}=\frac{\partial_jh_i(u)}{h_j(u)},\qquad i\ne j,
\end{equation}
which is known under the name the Darboux-Egoroff system:
\begin{equation}
\label{f03}
\begin{aligned}
\gamma_{ij}(u)&=\gamma_{ji}(u),\\
\partial_k\gamma_{ij}(u)&=\gamma_{ik}(u)\gamma_{kj}(u),\\
\sum_{k=1}^n\partial_k\gamma_{ij}(u)&=0,
\end{aligned}
\end{equation}
{}From (\ref{f01a}) and (\ref{f02}), we see that the Lam\'e coefficients $h_i$
satisfy:
\begin{equation}
\label{f08}
\partial_jh_i(u)=\gamma_{ij}(u)h_j(u),\quad i\ne j,\qquad
\partial_i h_i(u)=-\sum_{j\ne i}\gamma_{ij}(u)h_j(u).
\end{equation}

The flat coordinates $t^1,\dots,t^n$ of this metric can be found from
the linear system
\begin{equation}
\label{f06}
\partial_i\partial_j t^k=\Gamma_{ij}^i \partial_i t^k
+ \Gamma_{ji}^j \partial_j t^k ,
\quad i\ne j;\qquad\quad
\partial_i\partial_i t^k=\sum_{j=1}^n \Gamma_{ii}^j \partial_j t^k ,
\end{equation}
where $\Gamma_{ij}^k$ are Christoffel symbols:
\begin{equation}
\label{f09}
\Gamma_{ij}^i=\frac{\partial_j h_i}{h_i},\qquad
\Gamma_{ii}^j=(2\delta_{ij}-1)\frac{h_i\partial_j h_i}{h^2_j}.
\end{equation}

R. Martini and the author constructed in \cite{LM} solutions of the
Darboux-Egoroff system (\ref{f03}). These solutions were related to certain
points in the $SL_n$ Loop group orbit of the highest weight vector of the
homogeneous realization of the basic representation and were related to a
reduction of the $n$-component KP hierarchy. They had no real representation
theoretical explanation for these particular solution. In this paper we show
that all elements in the twisted Loop group orbit of $GL_n$ lead to solutions
of the Darboux-Egoroff system. The solutions of \cite{LM} are certain
homogeneous solutions in this orbit. This makes it possible to construct
non-homogeneous WDVV prepotentials $F$. Inspired by the papers \cite{Kr} and
\cite{AKrV},
we construct in section 5 for all elements in the twisted Loop group orbit
besides solutions of the Darboux-Egoroff system, also the related (deformed)
flat coordinates and the
WDVV prepotential $F$.

\section{Groups and Grassmannians}

Consider the space $H_n$ of Laurent series in $t$ (this $t$ has nothing to do
with the flat coordinates appearing in the previous section) with coefficients
in $\mathbb{C}^n$
\[
H_n=\{\sum_j c_jt^j | c_j\in\mathbb{C}^n,\ c_j=0\ \text{for }j<<0\}.
\]
Let $e_j$, $1\le j\le n$,  be a basis of $\mathbb{C}^n$,
Define
\begin{equation}
\label{f1}
v_{-n(k+1)+j-\frac{1}{2}}=v^{(j)}_{-k-\frac{1}{2}}:=e_jt^k,\qquad 1\le j\le n,\
 k\in\mathbb{Z},
\end{equation}
then an element of $H_n$ is a unique linear combination of, possibly infintely
many, $v_\ell$'s or equivalently $v^{(j)}_\ell$'s.
The space $H_n$ has a natural filtration ($j\in \mathbb{Z})$
\[
\dots H_n^{(j-1)} \subset H_n^{(j)}\subset H_n^{(j+1)}
\subset\dots,
\]
where
\begin{equation}
\label{f2}
H_n^{(j)}=\{\sum_k b_kv_k | b_k\in\mathbb{C},\ b_k=0\ \text{for }k>j\}.
\end{equation}
Let $H_n^*$
denote the space of linear functions $f$ on $H_n$ such that $
f(H_n^{(j)})=0$ for $j<<0$. On the direct sum
$\overline H_n=H_n\oplus H_n^*$ of these two spaces, one has a natural
symmetric nondegenerate bilinear form $(\cdot,\cdot)$ for which the spaces
 $H_n$ and $H_n^*$ are isotropic, it is defined by
$(f,v)=f(v)$ for  $v\in H_n$ and   $f\in H_n^*$.
Define "dual basis" elements to the elements defined in (\ref{f1}) as follows
\begin{equation}
\label{f3}
v_{-nk-j+\frac{1}{2}}^*=v_{-k-\frac{1}{2}}^{(j)*}=e_j^*t^k,
\end{equation}
then the bilinear form is given by the following formula's
\begin{equation}
\label{f4}
(e_i^*t^\ell ,e_jt^k)=\delta_{k+\ell,-1}\delta_{ij},\quad
(v_{r}^{(i)*},v_{s}^{(j)})=\delta_{r,-s}\delta_{ij},\quad
(v_r^*,v_s)=\delta_{r,-s}.
\end{equation}
Let $\mathbb{C}^{n*}=\oplus_{j=1}^n\mathbb{C}e_j^*$ then clearly
\[
H_n^*=\{\sum_j c_j^*t^j | c_j^*\in\mathbb{C}^{n*},\ c_j^*=0\ \text{for }j<<0\}.
\]
In analogy with (\ref{f2}), one also has the subspaces
\begin{equation}
\label{f5}
 H_n^{(j)^*}=\{\sum_k b_kv_k^* | b_k\in\mathbb{C},\ b_k=0\ \text{for }k>j\}.
\end{equation}
Let $(\cdot |\cdot )$ be the symmetric bilinear form on
$\mathbb{C}^n\oplus \mathbb{C}^{n*}$ for which
$\mathbb{C}^n$ and $ \mathbb{C}^{n*}$ are isotropic and
$(e_i^*|e_j)=\delta_{ij}$. Let $v(t)\in H_n$ and $f(t) \in H^*_n$, then
(\ref{f4}) is equivalent to
\begin{equation}
\label{f4a}
 (f(t),v(t))=\text{Res}_t (f(t)|v(t)),
\end{equation}
where $\text{Res}_t \sum a_it^i=a_{-1}$. Notice that the left hand side of
(\ref{f4a}) makes sense.

Now, notice that the space $\overline H_n^{(j)}:=H_n^{(j)}\oplus H_n^{(-j)^*}$
is a maximal isotropic subspace of $\overline H_n$ with respect to our bilinear
form $(\cdot,\cdot)$. Having these specific isotropic spaces in mind,
we want to define more general maximal isotropic subspaces. Let $\overline
V\subset\overline H_n$ be a  maximal isotropic linear subspace, which satisfies
the
following conditions:
\begin{equation}
\label{f6}
\overline V=V\oplus V^*\ \text{with }V\subset H_n,\ V*\subset H_n^*
\end{equation}
and
\begin{equation}
\label{f7}
 H_n^{(j)}\subset V,\ H_n^{(j)^*}\subset V^*\ \text{for }
j<<0.
\end{equation}
All such subspaces form the points of an infinite (isotropic) Grassmannian
$\overline {Gr}$. Clearly, since (\ref{f6}) holds, a $\overline V\in\overline
{Gr}$
induces two unique spaces $V$ and $V^*$, which can be separately regarded as
points of two Grassmannians
\[
Gr=\{V\subset H_n |  H_n^{(j)}\subset V, \ \text{for }
j<<0\}\ \text{and }
Gr^*=\{V^*\subset H_n^* |  H_n^{(j)*}\subset V*, \ \text{for }
j<<0\}.
\]
It is obvious that the converse also holds, i.e., if $V\in Gr$ (or $V^*\in
Gr^*$), then there exists a unique maximal space $V^*\in Gr^*$ (resp. $V\in
Gr$) such that $(V^*,V)=0$ and hence $\overline V=V\oplus V^*$ is a unique
point of $\overline {Gr}$.

The space $Gr$ is Sato's polynomial Grassmannian \cite{S}. Here it is coupled
to $Gr^*$ and $\overline{Gr}$. The reason why we introduce the latter two
spaces will become clear later on. It is well-known that the space
$Gr=\cup_{m\in\mathbb{Z}} Gr_m$, disjoint union of the spaces
\[
Gr_m=
\{V\in Gr|H_n^{(j)}\subset V\ \text{and dim }V/H_n^{(j)}=m-j\ \text{for }j<<0.
\}
\]
If $V\in Gr_m$, then $V^*$ belongs to
\[
Gr_{-m}^*=
\{V^*\in Gr|H_n^{(j)*}\subset V^*\ \text{and dim }V^*/H_n^{(j)*}=-m-j\
\text{for }j<<0.
\}
\]
In this situation, $\overline V=V\oplus V^*$ is an element of
\[
Gr_m=
\{\overline V=V\oplus V^*\in \overline{Gr}|V\in Gr_m\ \text{and }V^*\in
Gr^*_{-m}\}.
\]

Suppose that the space $V$ is invariant under multiplication with $t$, i.e.,
$V$ is an element of the restricted Grassmannian
\[
gr:=\{ V\in Gr|tV\subset V\},
\]
then it is clear from the above construction that its "dual space" $V^*$ is
unique. Now let $f(t)\in V^*$, then for all $v(t)\in V)$ we have
$\text{Res}_t (f(t)|v(t))=0$. Since $tv(t)\in V$, also
\[
\text{Res}_t(tf(t)|v(t))=
\text{Res}_t(f(t)|tv(t))=0\ \text{for all }v(t)\in V,
\]
hence $tf(t)\in V^*$. This means that also $V^*$ and $\overline V=V\oplus V^*$
satisfy $tV^*\subset V^*$, $t\overline V\subset\overline V$ and that it makes
sense to define
\[
gr^*:=\{V^*\in Gr^*|tV^*\subset V^*\} \qquad \text{and }\quad
\overline{gr}:=\{\overline V\in \overline{Gr}|
t\overline V\subset\overline V\}
\]
and
\[
gr_m:=gr\cap Gr_m,\quad  gr^*_m:=gr^*\cap Gr^*_m\quad\text{and }\
\overline{gr}_m:=\overline{gr}\cap\overline{Gr}_m.
\]

Consider $H_n^{(j)}$, $j\in \mathbb{Z}$, to be the  a fundamental system of
neighborhoods of zero, then $H_n$ becomes a topological vector space.
let $\overline a_\infty$ be the algebra of all continuous endomorphisms of
$H_n$. If one considers an elmenent of $H_n$ as an infinite vector with respect
to the basis $v_i$, $i\in \frac{1}{2}+\mathbb{Z}$,
then (see \cite{KP2})
\begin{equation}
\label{f8}
\overline a_\infty=\{(a_ij)_{i,j\in\frac{1}{2}+\mathbb{Z}}|
\text{for each }k\ \text{the number of non-zero }a_{ij}\ \text{with }i\le k\
\text{and }j\ge k\ \text{is finite}\}.
\end{equation}
Denote by $\overline A_\infty$ the group of invertible elements of the
associative algebra $\overline a_\infty$. Then  $\overline A_\infty$ acts
trasitively on $Gr$.
We can extend an element $g\in \overline A_\infty$ to an element, which by
abuse of notation we denote by the same letter, g of $O(\overline V)$, the
orthogonal group of $\overline V$ (see \cite{KL2}):
\begin{lemma}
  \label{l1}
Let $g\in \overline A_\infty$ be such that
\[
g v_j = \sum_{i\in\mathbb{Z}+\frac{1}{2}} A_{ij} v_i
\]
and let $A=(A_{ij})_{i,j \in
\mathbb{Z}+\frac{1}{2}
}$ and
$A^{-1} = (B_{ij})_{i,j\in\mathbb{Z}+\frac{1}{2}
}$. Then
\[
g v^*_j = \sum_{i\in\mathbb{Z}+\frac{1}{2}
} B_{-ji} v^*_{-i}.
\]

\end{lemma}

{\bf Proof.}
Suppose that $g v_{\ell}^* = \sum_k B_{-\ell , k} v_{-k}^*$, then it
follows from
\[
\begin{array}[h]{lcl}
\delta_{j, -\ell} = (v_j , v_{\ell}^*) &=& (\sum_i A_{ij} v_i ,
B_{-\ell k} v_{-k}^*)\\
&=& \sum_i A_{ij} B_{-\ell , i}
\end{array}
\]
that $B=A^{-1}$. \hfill{$\square$}
\vskip 10pt
\noindent
This lemma defines the (transitive) action of $\overline A_\infty$
on $Gr^*$ and hence on $\overline{Gr}$.

Let $L=\mathbb{C}[t,t^{-1}]$ be the algebra of Laurent polynomials in $t$.
The identification (\ref{f1}) gives us an embedding
\begin{equation}
\label{f9}
\phi:\text{Mat}_n(L) \to\overline a_\infty, \qquad
\phi(e_{ij}t^\ell)=\sum_{k\in\mathbb{Z}}
E_{nk+i-\frac{1}{2},n(k+\ell)+j-\frac{1}{2}},
\end{equation}
where $e_{ij}=(\delta_{ir}\delta_{js})_{1\le r,s\le
n}\in\text{Mat}_n(\mathbb{C})$
and $E_{ij}=(\delta_{ir}\delta_{js})_{r,s\in\frac{1}{2}+\mathbb{Z}}$.
This embedding gives rise to the embedding of the loop Lie algebra
$gl_n(L)$ into $\overline a_\infty$ and the loop group $GL_n(L)$ in
$\overline A_\infty$.
Multiplication with $t$, i.e. with $\sum_{i=1}^n e_{ii}t$,  commutes with the
action of $GL_n(L)$ and thus $GL_n(L)$ acts on $gr$. This action is transitive,
see \cite{PS}.
Moreover,  $\phi(t)=\phi(\sum_{i=1}^n
e_{ii}t)=\sum_{k\in\frac{1}{2}+\mathbb{Z}}E_{k,k+n}$.


Consider $H_n^{(j)}$, it has as a "natural basis" the elements $v_i$ with
$i<j$.
Its dual space $H_n^{(-j)*}$ has $v_i^*$ with $i<-j$ as "basis".
To these spaces we associate  special vectors in two semi-infinite wedge spaces
(see \cite{KP2}, \cite{KV})
\begin{equation}
\label{f10}
\begin{aligned}
|H_n^{(j)}\rangle&=v_{j-\frac{1}{2}}\wedge v_{j-\frac{3}{2}}\wedge
v_{j-\frac{5}{2}}
\wedge v_{j-\frac{7}{2}}\wedge\cdots,\\
\langle H_n^{(-j)^*}|&=\cdots\wedge v^*_{-j-\frac{7}{2}}\wedge
v^*_{-j-\frac{5}{2}}\wedge v^*_{-j-\frac{3}{2}}
\wedge v^*_{-j-\frac{1}{2}}.
\end{aligned}
\end{equation}
In fact we can associate to any element $W\in Gr_j$ and its dual space $W^*\in
Gr_{-j}^*$ elements in these wedge spaces, viz., we know that there exists an
$m<<0$ such that
 $H_n^{(m)}\subset W$ and $H_n^{(m)*}\subset W^*$. Then let
$w_{j-\frac{1}{2}},\ w_{j-\frac{3}{2}},\cdots, \ w_{m+\frac{1}{2}}$ be a basis
of $W\text{mod} H_n^{(m)}$ and  $w^*_{-j-\frac{1}{2}},\
w^*_{-j-\frac{3}{2}},\cdots, \ w^*_{m+\frac{1}{2}}$ be a basis of
$W^*\text{mod} H_n^{(m)*}$ then we put
\begin{equation}
\label{f11}
\begin{aligned}
|W\rangle&=w_{j-\frac{1}{2}}\wedge  w_{j-\frac{3}{2}}\wedge \cdots \wedge
w_{m+\frac{1}{2}}\wedge v_{m-\frac{1}{2}}\wedge v_{m-\frac{3}{2}}\wedge
v_{m-\frac{5}{2}}\wedge \cdots,\\
\langle W^*|&=\cdots\wedge v^*_{m-\frac{5}{2}}\wedge v^*_{m-\frac{3}{2}}\wedge
v^*_{m-\frac{1}{2}}\wedge w^*_{m+\frac{1}{2}}\wedge  w^*_{m+\frac{3}{2}}\wedge
\cdots \wedge w^*_{-j-\frac{1}{2}}.
\end{aligned}
\end{equation}
It is clear that upto a constant both
$|W\rangle$ and $\langle W^*|$ are independent of the choice of basis of
 $W\text{mod} H_n^{(m)}$ and
$W^*\text{mod} H_n^{(m)*}$. This gives us a map $\mu$  from $Gr$ and $Gr^*$
into $\mathbb{P}F$ and  $\mathbb{P}F^{*}$, where
$F=\oplus_{j\in\mathbb{Z}}F^{(j)}$, $F^*=\oplus_{j\in\mathbb{Z}}F^{(j)*}$ and
the spaces
$F^{(j)}$ and  $F^{(-j)*}$ are the vector spaces generated by the elements
on the right-hand side of (\ref{f11}) (see \cite{KP2},\cite{KV} or \cite{KR}
for more details).
Since $\overline A_\infty$ and its subgroup $GL_n(L)$ act on $Gr$ and $Gr^*$
we obtain a projective representation of these groups on $\mathbb{P}F$ and
$\mathbb{P}F^{*}$.

Let $\omega$ be the following orthogonal transformation of $\overline V$
with respect to  the bilinear form (\ref{f4}):
\begin{equation}
\label{f13}
\omega(v_m^{(a)})=(-)^{m+\frac{1}{2}}iv_m^{(a)^*},\qquad
\omega(v_m^{(a)*})=(-)^{m+\frac{1}{2}}iv_m^{(a)},
\qquad m\in\frac{1}{2}+\mathbb{Z},\ 1\le a\le n,
\end{equation}
then  $\omega (\overline H_n^{(0)})=
\overline H_n^{(0)}$ and
 $\omega (\overline {Gr}_m)=\overline{Gr}_{-m}$.
If $g\in\overline A_\infty$ is such that
\begin{equation}
\label{f13a}
gv_j^{(b)}=\sum_{a,k}A_{kj}^{(ab)}v_k^{(a)}\quad \text{then }
gv_j^{(b)*}=\sum_{k,a}
B_{-j,k}^{(b,a)} v_{-k}^{(a)*}\quad \text{with }\sum_{j,b
}A_{kj}^{(ab)}B_{j\ell}^{bc}=
\delta_{k\ell}\delta_{ab},
\end{equation}
then
\[
\begin{aligned}
\omega(gv_j^{(b)})=&\omega(g)\omega(v_j^{(b)})=\omega(g)(-)^{j+\frac{1}{2}}i
v_j^{(b)*}\\[2mm]
=&\sum_{a,k}A_{kj}^{(ab)}\omega(v_k^{(a)})=\sum_{a,k}A_{kj}^{(ab)}
(-)^{k+\frac{1}{2}}iv_k^{(a)*}.
\end{aligned}
\]
This and a similar calculation for $gv_j^{(b)*}$ gives
\begin{equation}
\label{f14}
\omega(g)v_j^{(b)}=\sum_{a,k}(-)^{k-j}B_{-j,-k}^{(ba)}v_k^{(a)},\qquad
\omega(g)v_j^{(b)*}=\sum_{a,k}(-)^{k-j}A_{kj}^{(ab)}v_k^{(a)*},
\end{equation}
and hence induces on  $GL_n(L)$ the automorphism:
\begin{equation}
\label{f15}
\omega (A(t))=((A(-t))^T)^{-1},\qquad A(t)\in\text{Mat}_n(L),
\end{equation}
where $A^T$ stand for the transposed of the matrix $A$.
The fixed point set of this automorphism is the twisted loop group
\[
GL_n(L)^{(2)}=\{ A(t)\in GL_n(L)|A(-t)=((A(t))^T)^{-1}\}.
\]
All this suggests to define a new skew-
symmetric bilinear form
$\langle\cdot,\cdot\rangle$ on $H_n$
\[
\langle u, v\rangle=(u,\omega(v)) \qquad u,v\in H_n,
\]
or in other terms cf. (\ref{f6})
\begin{equation}
\label{f16}
\langle u(t),v(t)\rangle= i \text{Res}_{t}(u(t)|v(-t))\qquad u(t),v(t)\in H_n,
\end{equation}
and new Grassmannians
\[
Gr^{(2)}=\{ W\in Gr_0|\langle u,v\rangle=0\ \text{for all }u,v\in W\},
\qquad gr^{(2)}= gr\cap Gr^{(2)}.
\]
We now want to show that $gr^{(2)}$ the homogeneous space is of the twisted
loop group $GL_n(L)^{(2)}$.
It is obvious, from the above discussion, that $g\cdot H_n^{(0)}\in gr^{(2)}$
for
any $g\in GL_n(L)^{(2)}$. To prove the converse, we will use the following
Theorem of \cite{PS}:
\begin{theorem}
\label{tf1}
Any loop in $g(t)\in Gl_n(L)$ can be uniquely factorized as $g(t)=g_u(t)\cdot
g_+(t)$,
with $g_+(t)\in GL_n(\mathbb{C}[t])$ and
$
g_u\in \Omega U_n=\{h(t)\in U_n(L)|h(1)=I_n\}
$.
Moreover,
$gr=\Omega U_n\cdot H_n^{(0)}$ and isotropy group of $H_n^{(0)}$ is the group
$U_n$ of constant loops.
\end{theorem}
Now let $W\in gr^{(2)}$, then we can write $W=g(t)\cdot H_n^{(0)}$ for certain
$g(t)\in\Omega U_n$. Clearly also $\omega(g(t))\cdot H_n^{(0)}=W$,
hence $(g(t))^{-1}\omega(g(t))=u\in U_n$. Since $\omega$ is an involution on
$GL_n(L)$ we find that $u\omega(u)=1$, and since $u$ is unitairy we deduce
that the following conditions hold for u:
\[
u=u^T\qquad\text{and } u\overline u^T=I_n.
\]
So we can find a real orthogonal matrix $v$ such that $u=vdv^T$, with $d$ a
diagonal matrix. Let $s$ be a diagonal matrix that satisfies $s^2=d$, then
$w=w^T=vsv^T$ is a unitary matrix that satisfies $u=w^2$ and
\[
\omega(g(t)w)=g(t)u\omega(w)=g(t)w^2 (w^T)^{-1}=g(t)w^2w^{-1}=g(t)w.
\]
Since  also $W=g(t)\cdot H_n^{(0)}=g(t)w\cdot H_n^{(0)}$, we find that
$W\in GL_n(L)^{(2)}\cdot H_n^{(0)}$ and we have proven the following
\begin{theorem}
\label{tf2}
\[
gr^{(2)}=GL_n(L)^{(2)}\cdot H_n^{(0)}.
\]
\end{theorem}

\section{The Clifford Algebra and Tau-Functions}

Using (\ref{f10}), we can associate to any point $W\in Gr_j$ and the
 corresponding $W^*\in Gr^*_{-j}$ vectors
$|W\rangle\in \mathbb{P}F^{(j)}$, $\langle W^*|\in \mathbb{P}F^{(-j)*}$
and hence upto a constants unique vectors
$|W\rangle\in F^{(j)}$ and  $\langle W^*|\in F^{(-j)*}$.
Clearly $\omega$ defined in (\ref{f13}) is not well defined on these vectors
in the spaces $F$ and $F^*$.
To solve this problem, we will define a Clifford algebra and its corresponding
spin module and we will fix $\omega$ on one of the vectors of the spin module.
In this construction we will only consider the space $F$, since the space $W^*$
corresponding to a $W\in Gr$ is always unique.

Recall that on the infinite space $\overline H_n=H_n\oplus H_n^*$
we have a symmetric bilinear form $(\cdot ,\cdot )$  given by (\ref{f4}).
Let $C\ell(\overline H_n)$ be the Clifford algebra on this space, i.e. the
associative algebra over $\mathbb{C}$ with unity $1$ wich has the folowing
defining relations
\begin{equation}
\label{f2.1}
uv+vu=(u,v)1,\qquad u,v\in \overline H_n.
\end{equation}
We obtain a obtain a representation $\psi$ of the clifford algebra on the space
$F$ by defining it on wedges as follows ($w, w_j\in H_n\ w^*\in H_n^*$ with
$w_j=v_{-j-m-\frac{1}{2}}$ for certain $m\in\mathbb{Z}$ and  all $j>>0$)
\begin{equation}
\label{f2.2}
\begin{aligned}
\psi(w)(w_{0}\wedge  w_{-1}&\wedge w_{-2}\wedge \cdots)=
w\wedge w_{0}\wedge  w_{-1}\wedge w_{-2}\wedge \cdots,\\
\psi(w^*)(w_{0}\wedge  w_{-1}&\wedge \cdots)=
\sum_{i=0}^\infty (-)^j (w^*,w_j)  w_{0}\wedge  w_{-1}\wedge\cdots
\wedge w_{j-1}\wedge w_{j+1}\wedge\cdots.
\end{aligned}
\end{equation}
The space $F$ is  the spin module for this Clifford algebra.
Let
\[
|0\rangle= v_{-\frac{1}{2}}\wedge v_{-\frac{3}{2}}\wedge v_{-\frac{5}{2}}\wedge
v_{-\frac{7}{2}}\wedge\cdots,
\]
then $F$ is the unique module generated by
\begin{equation}
\label{f2.3}
\psi(v_k)|0\rangle =\psi(v_k^*)|0\rangle=0\qquad  \text{for }k<0.
\end{equation}
It is straightforward to check that (cf. (\ref{f11}))
\begin{equation}
\label{f2.4}
\begin{aligned}
w_{j-\frac{1}{2}}\wedge  w_{j-\frac{3}{2}}&\wedge \cdots \wedge
w_{m+\frac{1}{2}}\wedge v_{m-\frac{1}{2}}\wedge v_{m-\frac{3}{2}}\wedge
\cdots\\
=\psi(w_{j-\frac{1}{2}})&\psi(w_{j-\frac{3}{2}})\cdots\psi( w_{m+\frac{1}{2}})
\psi(v_{-m-\frac{1}{2}}^*)\psi(v_{-m-\frac{3}{2}}^*)
\cdots\psi(v_{\frac{3}{2}}^*)\psi(v_{\frac{1}{2}}^*)|0\rangle.
\end{aligned}
\end{equation}

Define the fermionic fields ($z\in\mathbb{C}^\times$)
\begin{equation}
\label{f2.5}
\begin{aligned}
\psi^{+(j)}(z)=\sum_{k\in\mathbb{Z}+\frac{1}{2}}\psi^{+(j)}_kz^{-k-\frac{1}{2}}
&:=\sum_{k\in\mathbb{Z}+\frac{1}{2}} \psi(v_k^{(j)})z^{k-\frac{1}{2}},\\
\psi^{-(j)}(z)=\sum_{k\in\mathbb{Z}+\frac{1}{2}}\psi^{-(j)}_kz^{-k-\frac{1}{2}}
&:=\sum_{k\in\mathbb{Z}+\frac{1}{2}} \psi(v_k^{(j)*})z^{k-\frac{1}{2}}
\end{aligned}
\end{equation}
and bosonic fields $(1 \leq j \leq n)$ by
\begin{equation}
\label{2.1.5}
\alpha^{(j)}(z)= \sum_{k \in {\Bbb Z}} \alpha^{(j)}_{k} z^{-k-1}
= :\psi^{+(j)}(z) \psi^{-(j)}(z):
\end{equation}
where $:\ :$ stands for the {\it normal ordered product} defined in
the usual way $(\lambda ,\mu = +$ or $-$):
\begin{equation}
\label{2.1.6}
:\psi^{\lambda (i)}_{k} \psi^{\mu (j)}_{\ell}: = \begin{cases} \psi^{
\lambda (i)}_{k}
\psi^{\mu (j)}_{\ell}\ &\text{if}\ \ell \ge k ,\\
-\psi^{\mu (j)}_{\ell} \psi^{\lambda (i)}_{k} &\text{if}\ \ell <k
.\end{cases}
\end{equation}
Notice that $\psi^{+(j)}(z)=\psi(\delta(z-t)e_j)$ and  $\psi^{-(j)}(z)=\psi(\delta(z-t)e_j^*)$, where 
$\delta(z-t)=z^{-1}\sum_{k\in\mathbb{Z}}\left(\frac{z}{t}\right)^k$.
One checks (using e.g. the Wick formula) that these bosonic operators
 satisfy the canonical commutation relation of the associative
oscillator algebra:
\begin{equation}
\label{2.1.9}
[\alpha^{(i)}_{k},\alpha^{(j)}_{\ell}] =
k\delta_{ij}\delta_{k,-\ell},
\end{equation}
and one has
\begin{equation}
\label{2.1.10}
\alpha^{(i)}_{k}|0\rangle = 0 \ \text{for}\ k \ge 0.
\end{equation}

In order to express the fermionic fields $\psi^{\pm (i)}(z)$ in terms of
the bosonic fields $\alpha^{(i)}(z)$, we need some additional operators
$Q_{i},\ i = 1,\ldots ,n$, on $F$.  These operators are uniquely defined by
the following conditions:
\begin{equation}
\label{2.1.12}
Q_{i}|0\rangle = \psi^{+(i)}_{-\frac{1}{2}} |0\rangle ,\ Q_{i}\psi^{\pm
(j)}_{k} = (-1)^{\delta_{ij}+1} \psi^{\pm
(j)}_{k\mp \delta_{ij}}Q_{i}.
\end{equation}
They satisfy the following commutation relations:
\begin{equation}
\label{2.1.13}
Q_{i}Q_{j} = -Q_{j}Q_{i}\ \text{if}\ i \neq j,\ [\alpha^{(i)}_{k},Q_{j}] =
\delta_{ij} \delta_{k0}Q_{j}.
\end{equation}
\begin{theorem}
\label{t2.1}  (\cite{DJKM1}, \cite{JM})
\begin{equation}
\label{2.1.14}
\psi^{\pm (i)}(z) = Q^{\pm 1}_{i}z^{\pm \alpha^{(i)}_{0}} \exp
(\mp \sum_{k < 0} \frac{1}{k} \alpha^{(i)}_{k}z^{-k})\exp(\mp
\sum_{k > 0} \frac{1}{k} \alpha^{(i)}_{k} z^{-k}).
\end{equation}
\end{theorem}
{\bf Proof}. See \cite{TV}.

The operators on the right-hand side of (\ref{2.1.14}) are called vertex
operators.  They made their first appearance in string theory (cf.
\cite{FK}).

We can describe now the $n$-component boson-fermion
correspondence.  Let ${\mathbb C}[x]$ be the space of polynomials in
indeterminates $x = \{ x^{(i)}_{k}\},\ k = 1,2,\ldots ,\ i =
1,2,\ldots ,n$.  Let $N$ be a lattice with a basis $\delta_{1},\ldots
,\delta_{n}$ over ${\mathbb Z}$ and the symmetric bilinear form
$(\delta_{i}|\delta_{j}) = \delta_{ij}$, where $\delta_{ij}$ is the
Kronecker symbol.  Let
\begin{equation}
\label{2.2.1}
\varepsilon_{ij} = \begin{cases} -1 &\text{if $i > j$} \\
1 &\text{if $i \leq j$.} \end{cases}
\end{equation}
Define a bimultiplicative function $\varepsilon :\ N \times N\to \{
\pm 1 \}$ by letting
\begin{equation}
\label{2.2.2}
\varepsilon (\delta_{i}, \delta_{j}) = \varepsilon_{ij}.
\end{equation}
Let $\delta = \delta_{1} + \ldots + \delta_{n},\ M = \{ \gamma \in
N|\ (\delta | \gamma ) = 0\}$, $\Delta = \{ \alpha_{ij} :=
\delta_{i}-\delta_{j}| i,j = 1,\ldots ,n,\ i \neq j \}$.  Of course
$M$ is the root lattice of $s\ell_{n}({\mathbb C})$, the set $\Delta$
being the root system.

Consider the vector space ${\mathbb C}[N]$ with basis $e^{\gamma}$,\
$\gamma \in L$, and the following twisted group algebra product:
\begin{equation}
\label{2.2.3}
e^{\alpha}e^{\beta} = \varepsilon (\alpha ,\beta)e^{\alpha +
\beta}.
\end{equation}
Let $B = {\mathbb C}[x] \otimes_{\mathbb C} {\mathbb C}[N]$ be the tensor
product of algebras.  Then the $n$-component boson-fermion
correspondence is the vector space isomorphism
\begin{equation}
\label{2.2.4}
\sigma :F \to B,\qquad \text{with }\sigma:F^{(m)}\to B^{(m)}
\end{equation}
given by
\begin{equation}
\label{2.2.5}
\sigma (\alpha^{(i_{1})}_{-m_{1}} \ldots
\alpha^{(i_{s})}_{-m_{s}}Q_1^{k_{1}}\ldots Q_n^{k_{n}}|0\rangle ) = m_{1}
\ldots
m_{s}x^{(i_{1})}_{m_{1}} \ldots x^{(i_{s})}_{m_{s}} \otimes
e^{k_{1}\delta_{1} + \ldots + k_{n}\delta_{n}} .
\end{equation}
The transported action of the operators $\alpha^{(i)}_{m}$ and $Q_{j}$ looks
as follows:
\begin{equation}\label{2.2.8}
\begin{cases}
\sigma \alpha^{(j)}_{-m}\sigma^{-1}(p(x) \otimes e^{\gamma}) =
mx^{(j)}_{m}p(x)\otimes e^{\gamma},\ \text{if}\ m > 0, &\  \\
\sigma \alpha^{(j)}_{m} \sigma^{-1}(p(x) \otimes e^{\gamma}) = \frac{\partial
p(x)}{\partial x_{m}} \otimes e^{\gamma},\ \text{if}\ m > 0, &\  \\
\sigma \alpha^{(j)}_{0} \sigma^{-1} (p(x) \otimes e^{\gamma}) =
(\delta_{j}|\gamma ) p(x) \otimes e^{\gamma} , &\ \\
\sigma Q_{j} \sigma^{-1} (p(x) \otimes e^{\gamma}) = \varepsilon
(\delta_{j},\gamma)  p(x) \otimes e^{\gamma + \delta_{j}}
. & \
\end{cases}
\end{equation}
The transported action of the fermionic fields is as follows:
\begin{equation}
\label{2.2.9}
\sigma\psi^{\pm(j)}(z)\sigma^{-1}=
e^{\pm \delta_j}z^{\pm \delta_j}\exp (\pm \sum^{\infty}_{k=1}
x^{(j)}_{k} )
\exp
( \mp \sum^{\infty}_{k=1}  \frac{\partial}{\partial
x^{(j)}_{k}} \frac{z^{-k}}{k})
\end{equation}

We will now determine the second part of the boson--fermion
correspondence, i.e., we want to determine $\sigma$ of the elements
(\ref{f2.4}).
Since for our purpose we are only interested in $Gr^{0)}$ we will assume that
this element is a wedge in $F^{(0)}$, i.e. let
\begin{equation}
\label{3.2}
\tau=A_{-\frac{1}{2}}\wedge
A_{-\frac{3}{2}}\wedge
A_{-\frac{5}{2}}\wedge
\cdots\in F^{(0)} \quad\text{with } A_{-p}=v_{-p} \ \text{for all } p>P>>0.
\end{equation}
To such an element we can associate an element in $A=(A_{ij})\in\overline
A_\infty$
such that $Av_{k}=A_{-k}$ for all $k>0$. Notice that
$A_{ij}=\delta_{ij}$ for $j<-P$.
Then R. Martini and the author showed in \cite{LM} the following
\begin{proposition}
\label{p1}
Let $\sigma(\tau)=\sum_{\alpha\in M}\tau_\alpha(x)e^\alpha$.
Assume that $\alpha=\sum_{j=1}^nk_j\delta_j$ and suppose that
\[
Q_1^{k_1}Q_2^{k_2}\cdots
Q_n^{k_n}|0\rangle=\lambda_\alpha v_{j_{-\frac{1}{2}}}\wedge
v_{j_{-\frac{3}{2}}}\wedge
v_{j_{-\frac{5}{2}}}\wedge\cdots,
\]
with
$j_{-\frac{1}{2}}>
j_{-\frac{3}{2}}>
j_{-\frac{5}{2}}\cdots$ and  $j_{-q}=-q$ for all $q>Q>>0$ and
$\lambda_\alpha=\pm 1$,
then
\[
\tau_\alpha(x)=\lambda_\alpha
\det\left(
\sum_{-R<\ell<0}
\sum_{r=j_{-\frac{1}{2}},
j_{-\frac{3}{2}}
,\cdots,
j_{-R+\frac{1}{2}}}
\sum_{\scriptsize
\begin{array}{c}
{1\le j\le n, q\in\mathbb{Z}+\frac{1}{2}}\\
{nq-\frac{1}{2}(n-2j+1)=r}
\end{array}
}
\left (\sum_{k=0}^\infty A_{r+nk,\ell}S_k(x^{(j)})\right )E_{r,\ell}\right
),
\]
where $R=\max (P,Q)$ and
$S_k(y)$ are the elementary Schur functions defined by
$\sum_{k\in\mathbb{Z}}S_k(y)z^k=\exp (\sum_{k=1}^\infty y_kz^k)$.
In particular if $1\le i<j\le n$ and $\alpha=0$, $\delta_i-\delta_j$,
$\delta_j-\delta_i$, respectively, then
$\lambda_0=1$, $\lambda_{\delta_i-\delta_j}=(-1)^{n-j}$,
$\lambda_{\delta_j-\delta_i}=(-1)^{n-i+1}$ and
$(j_{-\frac{1}{2}},
j_{-\frac{3}{2}}
,\cdots)=(-\frac{1}{2},-\frac{3}{2},-\frac{5}{2},\ldots)$,
$=(i-\frac{1}{2},-\frac{1}{2},-\frac{3}{2},\ldots,
j-n+\frac{1}{2},j-n+\frac{3}{2}\ldots)$,
$=(j-\frac{1}{2},-\frac{1}{2},-\frac{3}{2},\ldots,
i-n+\frac{1}{2},i-n+\frac{3}{2}\ldots)$, respectively.
\end{proposition}

We now want to know what happens if we apply $\omega$ to  such tau-functions.
Since $(\omega(u),\omega(v))=(u,v)$, $\omega$ extends to an automorphism of
order 4 on the Clifford algebra.
Next notice that $|0\rangle=|H_n^{(0)}\rangle\in \mathbb{P}F$. Since
$\omega(H_n^{(0)})=H_n^{(0)*}$, it makes sense to extend $\omega$ to $F$
by fixing it on $|0\rangle$ as $\omega(|0\rangle)=|0\rangle$.
It is then obvious that there is a one to one correspondence between elements
$w\in F^{(0)}$ that satisfy
$\omega(w)=\lambda w$ for certain $\lambda\in\mathbb{C}$ and
points
$W\in Gr^{(2)}$. Let $W\in Gr^{(2)}$ and
\[
w=w_{-\frac{1}{2}}\wedge w_{-\frac{3}{2}}\wedge \cdots\wedge
w_{-m+\frac{1}{2}}\wedge v_{-m-\frac{1}{2}}\wedge v_{-m-\frac{3}{2}}
\wedge\cdots\in F^{(0)}
\]
be the corresponding vector. Since $\langle w_i,w_j\rangle =\langle
w_i,v_k\rangle=0$, for all $0>i,j>-m$ and all $k<-m$
we obtain that all $w_i$ are of the form $w_i=\sum_{-m<j<m} w_{ji}v_j$.
Hence we can find vectors $w_{\frac{1}{2}}, w_{\frac{3}{2}},\ldots
w_{m-\frac{1}{2}}$, of the same form such that $\langle w_i, w_j\rangle=
\langle v_i,v_j\rangle$ for all $-m<i,j<m$. Thus $W =(w_{ij})_{-m<i,j<m}
$ is a Symplectic matrix and must have determinant equal to 1.
This makes it possible to write
$w$ in two ways, viz,
\begin{equation}
\label{f2.27}
w=\psi(w_{-\frac{1}{2}})\psi( w_{-\frac{3}{2}}) \cdots
\psi(w_{-m+\frac{1}{2}})\psi(v^*_{m-\frac{1}{2}})\psi( v^*_{m-\frac{3}{2}})
\cdots
\psi(v^*_{\frac{1}{2}})|0\rangle
\end{equation}
and
\begin{equation}
\label{f2.28}
\begin{aligned}
w=\psi(\omega(\langle
w_{\frac{1}{2}}&,w_{-\frac{1}{2}}\rangle^{-1}w_{-\frac{1}{2}}))
\psi(\omega(\langle w_{\frac{3}{2}},w_{-\frac{3}{2}}\rangle^{-1}
w_{-\frac{3}{2}}) )\cdots\\
\cdots &
\psi(\omega(\langle
w_{m-\frac{1}{2}},w_{-m+\frac{1}{2}}\rangle^{-1}w_{-m+\frac{1}{2}}))
\psi(v_{m-\frac{1}{2}})\psi( v_{m-\frac{3}{2}}) \cdots
\psi(v_{\frac{1}{2}})|0\rangle.
\end{aligned}
\end{equation}
It is then straightforward to check that $\omega(w)$ of the representation
(\ref{f2.27}) exactly gives (\ref{f2.28}).
This means that $\omega(w)=w$ for all elements $w\in F^{(0)}$ corresponding to
$W\in Gr^{(2)}$.

Next notice that
\[
\omega(\psi^{\pm(j)} (z))=\psi^{\mp(j)}(-z),
\]
and hence
\begin{equation}
\label{f2.23}
\omega (\alpha^{(j)}(z))=\alpha^{(j)}(-z),
\end{equation}
from which we deduce that
\[
\omega(\delta_j)=-\delta_j\qquad \text{and
}\omega(x_k^{(j)})=(-)^{k+1}x_k^{(j)}.
\]
Here we write, as an abuse of notation, $\omega$ for $\sigma\omega\sigma^{-1}$.
Next we want to calculate what $\omega$ does with $Q_j$.
Notice first,using (\ref{2.1.14}), that
\begin{equation}
\label{f2.24}
Q^{\pm 1}_{j}
=\exp
(\pm \sum_{k < 0} \frac{1}{k} \alpha^{(j)}_{k}z^{-k})
\psi^{\pm (j)}(z)
\exp(\pm\sum_{k > 0} \frac{1}{k} \alpha^{(j)}_{k} z^{-k})
z^{\mp \alpha^{(j)}_{0}}
\end{equation}
and that we may replace $z$ in this formula (\ref{f2.24}) by
$-z$, since the left-hand side is independent of $z$.
So,
\[
\begin{aligned}
\omega(Q^{\pm 1}_{j})&=i
\exp(\mp \sum_{k < 0} \frac{1}{k} \alpha^{(j)}_{k}(-z)^{-k})
\psi^{\mp (j)}(-z)
\exp(\mp\sum_{k > 0} \frac{1}{k} \alpha^{(j)}_{k} (-z)^{-k})
z^{\pm \alpha^{(j)}_{0}}\\
\ &=iQ^{\mp 1}_{j}(-)^{\alpha^{(j)}_{0}}.
\end{aligned}
\]
Thus we find for the operators $e^{\pm\delta_j}$:
\[
\omega(e^{\pm\delta_j})=ie^{\mp\delta_j}(-)^{\delta_j}.
\]
So we conclude that
\begin{equation}
\label{f2.25}
\begin{aligned}
\omega &\left (\tau_0(x_k^{(a)})+\sum_{1\le i< j\le n}
\left(
\tau_{\delta_i-\delta_j}(x_k^{(a)})e^{\delta_i-\delta_j}
+
\tau_{\delta_j-\delta_i}(x_k^{(a)})e^{\delta_j-\delta_i}
\right ) +\cdots\right )\\
\ &\quad=\tau_0((-)^{k+1}x_k^{(a)})-\sum_{1\le i< j\le n}
\left(
\tau_{\delta_j-\delta_i}((-)^{k+1}x_k^{(a)})e^{\delta_i-\delta_j}
+
\tau_{\delta_i-\delta_j}((-)^{k+1}x_k^{(a)})e^{\delta_j-\delta_i}
\right )+\cdots.
\end{aligned}
\end{equation}

Next assume that $W\in gr_0$.
To this subspace corresponds an upto a multiple factor unique a vector $w\in
F^{(0)}$.
Since $\sum_{i=1}^n te_{ii}W\subset W$, we can find special  linearly
independent vectors $w_1, w_2,\ldots ,w_n\in W$
such that $t^\ell w_j=(\sum_{i=1}^n
te_{ii})^\ell w_j\in H_n^{(-\ell n)}$ for all $1\le j\le n$
 and such that
\[
w=w_1\wedge  w_2\wedge \cdots \wedge w_n\wedge
tw_1\wedge  tw_2\wedge \cdots  \wedge
 t^{\ell-1}w_n\wedge
v_{-\ell n-\frac{1}{2}}\wedge v_{-\ell n-\frac{3}{2}}\wedge\cdots.
\]
{}From this presentation of $w$ one easily sees that the action of
\[
\sum_{j=1}^n\alpha_k^{(j)}w=0\qquad\text{for all }\ k>0,
\]
This leads to
\begin{equation}
\label{f2.26}
\sum_{j=1}^n \frac{\partial \tau_\alpha(x)}{\partial x_k^{(j)}}=0
\qquad\text{for all }\ k>0
\end{equation}
and hence to the following
\begin{proposition}
\label{ft2}
Tau-functions $\tau_W(x)=\sum_{\alpha\in M}\tau_\alpha(x_k^{(a)})e^\alpha$
corresponding to $W\in gr^{(2)}$ satisfy the following conditions:
\[
\begin{aligned}
(1)&\ \sum_{j=1}^n \frac{\partial \tau_\alpha(x)}{\partial x_k^{(j)}}=0
\qquad\text{for all }\ k>0,\\
(2)&\ \tau_0((-)^{k+1}x_k^{(a)})=\tau_0(x_k^{(a)}),\\
(3)&\ \tau_{\delta_j-\delta_i}((-)^{k+1}x_k^{(a)})=
-\tau_{\delta_i-\delta_j}(x_k^{(a)})\qquad\text{for all }1\le i,j\le n,\ i\ne
j.
\end{aligned}
\]
\end{proposition}

\section{The KP hierarchy as a dynamical system}

It is well known,
see e.g. \cite{KP2}, that $\tau_W$ corresponds to a $W\in Gr_0$ if and only if
$\tau_W$ satisfies the the KP hierarchy, i.e., the following equation:
\begin{equation}
\label{2.3.1}
\text{Res}_{z=0} \sum^{n}_{j=1} \psi^{+(j)}(z)\tau
\otimes \psi^{-(j)}(z)\tau  = 0,\ \tau \in F^{(0)}.
\end{equation}
Using the boson-fermion correspondence we can write this equation as a family
of equation on certain $n\times n$ wave functions ($\alpha\in \text{supp
}\tau=\{\alpha \in M|\tau_\alpha\ne 0\}$)
\begin{equation}
\label{3.3.2}
V^{\pm} (\alpha,x,z) = (V^{\pm}_{ij}(\alpha ,x,z))^{n}_{i,j=1},
\end{equation}
(see \cite{KV} for more details) where
\begin{equation}
\label{3.3.3}
\begin{aligned}
\ &V^{\pm}_{ij}(\alpha ,x,z) :=
\varepsilon (\delta_{j} , \alpha + \delta_{i})
 z^{(\delta_{j}|\pm \alpha + \alpha_{ij})} \\
\ & \times \exp (\pm \sum^{\infty}_{k=1} x^{(j)}_{k} z^{k})
\exp(\mp \sum^{\infty}_{k=1} \frac{\partial}{\partial
x^{(j)}_{k}} \frac{z^{-k}}{k}) \tau_{\alpha  \pm
\alpha_{ij}}  (x)/\tau_{\alpha}(x).
\end{aligned}
\end{equation}
The equations are:
\begin{equation}
\label{3.3.4}
\text{Res}_{z=0}V^{+}(\alpha,x,z)V^{-}(\beta, x^{\prime},z)^T
= 0\
\text{for all}\ \alpha ,\beta \in \text{supp }\tau.
\end{equation}
Define $n \times n$ matrices $W^{\pm (m)} (\alpha ,x)$ by the
following generating series (cf. (\ref{3.3.3})):
\begin{equation}
\label{3.3.5}
\sum^{\infty}_{m=0}
W^{\pm (m)}_{ij} (\alpha ,x)(\pm z)^{-m}
= \varepsilon_{ji}z^{\delta_{ij}-1} (\exp \mp
\sum^{\infty}_{k=1} \frac{\partial}{\partial x^{(j)}_{k}}\frac{z^{-k}}{k})
\tau_{\alpha \pm
\alpha_{ij}} (x))/\tau_{\alpha} (x).
\end{equation}
Note that
\begin{equation}
\label{3.3.6}
W^{\pm (0)}(\alpha ,x) = I_{n},
\end{equation}
\begin{equation}
\label{3.3.7}
W^{\pm (1)}_{ij}(\alpha ,x) =
\begin{cases} \varepsilon_{ji}
\tau_{\alpha \pm \alpha_{ij}}/\tau_{\alpha} &\text{if}\ i \neq j \\
- \tau^{-1}_{\alpha} \frac{\partial \tau_{\alpha}}{\partial
x^{(i)}_{1}} &\text{if}\ i = j, \end{cases}
\end{equation}

We see from (\ref{3.3.3}) that $V^{\pm}(\alpha ,x,z)$ can be written in the
following form:
\begin{equation}
\label{3.3.9}
V^{\pm}(\alpha ,x,z) = \sum^{\infty}_{m=0}
W^{\pm (m)}(\alpha ,x)(\pm
z)^{-m}R^{\pm}(\alpha ,\pm z)S^\pm(x,z),
\end{equation}
where
\begin{equation}
\label{3.3.10}
\begin{aligned}
R^{\pm}(\alpha ,z) &= \sum^{n}_{i=1}
\varepsilon (\delta_{i}, \alpha ) E_{ii} (\pm z)^{\pm
(\delta_{i}|\alpha )},\\
S^{\pm}(x,z) &= \sum^{n}_{i=1} e^{\pm\sum_{j=1}^\infty  x_j^{(i)}
z^j}E_{ii}.
\end{aligned}
\end{equation}
Here  $E_{ij}$ stands for the $n \times n$ matrix whose
$(i,j)$ entry is $1$ and all other entries are zero.  Now let
$\partial=\sum_{j=1}^n \frac{\partial}{\partial x_1^{(j)}}$, then
$V^{\pm}(\alpha ,x,z)$ can be
written in
terms of formal pseudo-differential operators (see \cite{KV} for more details).
Let
\begin{equation}
\label{3.3.11}
P^{\pm}(\alpha ) \equiv P^{\pm} (\alpha ,x,\partial ) =
I_{n} + \sum^{\infty}_{m=1} W^{\pm (m)} (\alpha ,x)\partial^{-m},\
R^{\pm}(\alpha ) = R^{\pm}(\alpha ,\partial),
\end{equation}
then
\begin{equation}
\label{3.3.12}
V^{\pm}(\alpha ,x,z)
=P^{\pm } (\alpha )
R^{\pm}(\alpha)S^\pm (x,z)
\end{equation}
 and one can prove that $P^-(\alpha)=P^+(\alpha)^{*-1}$ and the following
Lemma:

\begin{proposition}
\label{l3.4}  Let $\alpha,\beta\in \text{supp }\tau$, then
$P^{+}(\alpha)$ satisfies the Sato equations:
\begin{equation}
\label{3.4.2}
\frac{\partial P^+(\alpha)}{\partial x^{(j)}_{k}} = -(P^+(\alpha)E_{jj}
\partial^{k}  P^+(\alpha)^{-1})_{-}  P^+(\alpha)
\end{equation}
and $P^+(\alpha), \ P^+(\beta)$ satisfy
\begin{equation}
\label{3.3.16}
(P^{+}(\alpha)R^{+}(\alpha - \beta)P^{+}(\beta)^{-1})_{-} = 0\
\text{for all}\ \alpha ,\beta \in \text{supp}\ \tau .
\end{equation}
\end{proposition}
This is another formulation of the $n$-component KP hierarchy (see \cite{KV}).
Introduce the following formal pseudo-differential
operators $L(\alpha),\  C^{(j)}(\alpha )$:
\begin{equation}
\label{3.4.1}
\begin{aligned}
L(\alpha ) \equiv L(\alpha,x,\partial)
  & = P^{+}(\alpha)  \partial  P^{+}(\alpha)^{-1}, \\
C^{(j)}(\alpha) \equiv C^{(j)}(\alpha ,x,\partial) &=
P^{+}(\alpha)E_{jj} P^{+}(\alpha)^{-1}, \\
\end{aligned}
\end{equation}
then related to the Sato equation is the following linear system
\begin{equation}
\label{3.5.5}
\begin{aligned}
L(\alpha)V^{+}(\alpha ,x,z) &= zV^{+}(\alpha ,x,z) ,\\
C^{(i)}(\alpha)V^{+}(\alpha ,x,z) &=V^{+}(\alpha ,x,z) E_{ii} , \\
\frac{\partial V^{+}(\alpha ,x,z) }{\partial x^{(i)}_{k}}
&=(L(\alpha)^kC^{(i)}(\alpha))_+ V^{+}(\alpha ,x,z) .\\
\end{aligned}
\end{equation}

To end this section we write down explicitly some of the Sato
equations (\ref{3.4.2}) on the matrix elements $W^{(s)}_{ij}$ of the
coefficients $W^{(s)}(x)$ of the pseudo-differential operator
\[
P =P^+(\alpha)
= I_{n} + \sum^{\infty}_{m=1} W^{(m)}(x) \partial^{-m}.
\]
We shall write $W=W^{(1)}$ and $W_{ij}$ for $W^{(1)}_{ij}$ to simplify
notation,
then the simplest Sato equation is
\begin{equation}
\label{3.17a}
\frac{\partial P}{\partial x^{(k)}_{1}}=[\partial E_{kk},P]+[W,E_{kk}]P.
\end{equation}
In particular we
have for $i \neq k$:
\begin{equation}
\label{3.7.1}
\frac{\partial W_{ij}}{\partial x^{(k)}_{1}} = W_{ik} W_{kj} -
\delta_{jk} W^{(2)}_{ij}.
\end{equation}
The equation (\ref{3.17a}) is equivalent to the following equation for
$V=V^+(\alpha)$:
\begin{equation}
\label{3.17b}
\frac{\partial V}{\partial x^{(k)}_{1}}=(E_{kk}\partial +[W,E_{kk}])V.
\end{equation}

\section{The Darboux-Egoroff system}
Define
\begin{equation}
\label{f4.1}
w_{ij}(x)=W_{ij}^{(1)}(0,x),
\end{equation}
then from the previous section we know that
$w_{ij}(x)$ satisfies
\begin{equation}
\label{f4.2}
\frac{\partial w_{ij}(x)}{\partial
x_1^{(k)}}=w_{ik}(x)w_{kj}(x)\quad i\ne k\ne j.
\end{equation}
If we moreover assume that the wave function corresponds to a point $W\in
gr^{(2)}$, we also have
\begin{equation}
\label{f4.3}
\sum_{k=1}^n\frac{\partial w_{ij}(x)}{\partial
x_1^{(k)}}=0,
\end{equation}
and
\begin{equation}
\label{f4.4}
w_{ij}((-)^{k+1}x_k^{(a)})=w_{ji}(x_k^{(a)}).
\end{equation}
This makes it possible to obtain solutions of the Darboux-Egoroff system, viz
define
\begin{equation}
\label{4.1}
\gamma_{ij}(x)=w_{ij}(x)|_{x_{2k}^{(\ell)}=0\ \text{for all }k\ge 1,\ 1\le \ell\le
n},
\end{equation}
then these $\gamma_{ij}$ satisfy the equations (\ref{f03}).
Thus we have obtained the main theorem of this paper.
\begin{theorem}
\label{ft4}
Let $W\in gr^{(2)}=GL_n(L)^{(2)}\cdot H_N^{(0)}$ and let
$\tau(x)=\sum_{\alpha\in M}\tau_\alpha(x)e^\alpha$ be the corresponding
tau-function. Then the
\[
\gamma_{ij}(x)= \epsilon_{ji}
\left (\frac{\tau_{\delta_i-\delta_j}(x)}{\tau_0(x)}\right )_{x_{2k}^{(\ell)}=0
\ \text{for all }k\ge 1,\ 1\le \ell\le n}
\]
are solutions of the Darboux-Egoroff system (\ref{f03}) for $u_i=x_1^{(i)}$.
\end{theorem}
It is obvious that one can construct even more tau-functions that correspond to
$W\in gr_0$ and wich lead to solutions of the Darboux-Egoroff system.
Namely, if we take a tau-function which comes from a $W\in gr^{(2)}$, then we
can always multiply it with an element $e^\beta$ for $\beta \in M$. The
$\tau_\beta(x)$ and the $\tau_{\beta+\delta_i-\delta_j}(x)$ of this new
tau-function also lead to solutions of the  Darboux-Egoroff system.

It is easy to see from theorem \ref{ft2} that the wave functions satisfy
\begin{equation}
\label{f5.1}
\sum_{i=1}^n \frac{\partial V^\pm (\alpha,x,z)}{\partial x_1^{(i)}}=zV^\pm
(\alpha,x,z)
\end{equation}
This means that we do not really have formal
pseudo-differential operators, but rather formal matrix-valued
Laurent series in $z^{-1}$. The Sato equation takes the following
simple form.
Let $P(z)=P^+(\alpha,x,z)=I+Wz^{-1}=\cdots$ then
\[
\frac{\partial P(z)}{\partial x^{(j)}_{k}} = -(P(z)E_{jj}
P(z)^{-1}z^k)_{-} P(z).
\]
and equation (\ref{3.17b}) turns into
\begin{equation}
\label{4.a}
\frac{\partial V^+(\alpha,x,z)}{\partial x^{(k)}_{1}}=(zE_{kk} +[W,E_{kk}])V^+(\alpha,x,z).
\end{equation}

Next let
\begin{equation}
\label{f5.2}
\Phi^\pm(x,z)=V^\pm(0,x,z)|_{x_{2k}^{(i)}=0\ \text{for all }k\ge 1,\ 1\le i\le
n},
\end{equation}
then it is straightforward to check that
\[
\Phi^-(x,z)=\Phi^+(x,-z).
\]
Thus
\[
\text{Res}_z\Phi^+(x,z)\Phi^+(x',-z)^T=0
\]
from which one deduces, when one takes $x=x'$, that
\[
\Phi^+(x,z)\Phi^+(x,-z)^T=I_n.
\]
Let $\Gamma(x)=(\gamma_{ij}(x))_{1\le i,j\le n}$, then $\Phi(x,z):=\Phi^+(x,z)$
satisfies:

\begin{equation}
\label{f5.3}
\begin{aligned}
\Phi(x,z)\Phi(x,-z)^T&=I_n,\\
\sum_{j=1}^n \frac{\partial \Phi(x,z)}{\partial x^{(j)}_{1}}
&=z\Phi(x,z),\\
\frac{\partial \Phi(x,z)}{\partial x^{(k)}_{1}}&=(zE_{kk}
+[\Gamma(x),E_{kk}])\Phi(x,z),
\end{aligned}
\end{equation}
Following \cite{Du1}, we want solutions of this system for $z=0$. However, just
putting $z=0$ in (\ref{f5.3}) does not make sense. There is a way to construct
such solutions, viz. let $\tau=g(t)|0\rangle$, with $g(t)=\sum_i A(i)t^i\in
GL(L)^{(2)}$,
so in particular $g(-t)^T=g(t)^{-1}$,
then
\[
\psi(g(t)t^{-1}e_j)\tau\ne 0\ \text{ and }\psi(g(t)t^ke_j)\tau=0,\
\text{ for all }1\le j\le n,\ k\ge 0.
\]
Using the fermionic fields we can rewrite this to
\[
\text{Res}_z\sum_i\sum_{k=1}^n A(i)_{kj}z^{i-1}\psi^{+(k)}(z)\tau\ne 0
\ \text{ and }
\text{Res}_z\sum_i\sum_{k=1}^n A(i)_{kj}z^{i+\ell}\psi^{+(k)}(z)\tau=0
\]
for all $1\le j\le n$, $\ell\ge 0$ and thus
\[
\text{Res}_z z^{-1} V^+(0,x,z) g(z)\ne 0
\ \text{ and }
\text{Res}_z z^\ell V^+(0,x,z) g(z)=0\qquad\text{for all }\ell\ge 0.
\]
Now define
\[
\Psi(x,z):=z^{-1}\Phi(x,z)g(z),
\]
then this satisfies
\begin{equation}
\label{f5.4}
\begin{aligned}
\Psi(x,z)\Psi(x,-z)^T&=-z^{-2}I_n,\\
\sum_{j=1}^n \frac{\partial \Psi(x,z)}{\partial x^{(j)}_{1}}
&=z\Psi(x,z),\\
\frac{\partial \Psi(x,z)}{\partial x^{(k)}_{1}}&=(zE_{kk}
+[\Gamma(x),E_{kk}])\Psi(x,z),\\
\text{Res}_z z^\ell\Psi(x,z)&=0\qquad \text{for all }\ \ell>0.
\end{aligned}
\end{equation}
We thus get (c.f. \cite{Du1},\cite{Du2}):
\begin{proposition}
\label{p5.1}
Let $\Psi(x,z)$ be constructed as above.
Define
\[
\psi(x)=(\psi_{ij}(x))_{1\le i,j\le n}:=\text{Res}_z \Psi(x,z),
\]
Then these $\psi_{ij}$'s satisfy the equations
\begin{equation}
\label{ff5.1}
\frac{\partial\psi_{ij}}{\partial x_1^{(k)}}=
\gamma_{ik}\psi_{kj},\qquad k\ne i,\qquad
\sum_{k=1}^n
\frac{\partial\psi_{ij}}{\partial x_1^{(k)}}= 0.
\end{equation}
with
$\gamma_{ij}$ given by (\ref{4.1}) and the
formula's
\begin{equation}
\label{5.8}
\begin{aligned}
h_i&=\psi_{i1},\\
\eta_{\alpha\beta}&=\sum_{i=1}^n
\psi_{i\alpha}\psi_{i\beta}=\delta_{\alpha\beta},\\
\frac{\partial t^\alpha}{\partial x_1^{(i)}}&=\psi_{i1}\psi_{i\alpha},\\
c_{\alpha\beta}^\gamma=c_{\alpha\beta\gamma}&=\sum_{i=1}^n
\frac{\psi_{i\alpha}\psi_{i\beta}\psi_{i\gamma}}{\psi_{i1}},\\
\end{aligned}
\end{equation}
determine (locally) a semisimple Frobenius manifold on the
domain $x_1^{(i)}\ne x_1^{(j)}$ and $\psi_{11}\psi_{21}\cdots\psi_{n1}\ne 0$.
\end{proposition}

Define
\[
\begin{aligned}
\Theta(x,z)':=&z^2\sum_{i=1}^n\psi_{i1}E_{ii}\Psi(x,z),\\
\Theta(x,z)=&\left(\theta_1(x,z)\ \theta_2(x,z)\ \theta_3(x,z)\
\cdots\theta_n(x,z)\right):= z\left(\psi_{11}\ \psi_{21}\
\psi_{31}\cdots\psi_{n1}\right)
\Psi(x,z),
\end{aligned}
\]
then it is straightforward to check that
\begin{equation}
\label{5.8a}
\begin{aligned}
\text{Res}_z z^{k} \Theta(x,z)&=0,\qquad \text{Res}_z z^{k-1}
\Theta(x,z)'=0\quad \text{for all }k\ge 0,\\[2mm]
\Theta(x,z)'&=\left(\frac{\partial \theta_j(x,z)}{\partial
x_1^{(i)}}\right)_{1\le i,j\le n},\\[2mm]
\Theta(x,z)'\Theta(x,-z)'{}^{T}&=-z^2\sum_{i=1}^nh_i^2(x)E_{ii},\\[2mm]
z^{-1}\Theta(x,z)'\Theta(x,-z)^T&=\left( h_1^2(x)\ h_2^2(x)\ h_3^2(x)\ \cdots
h_n^2(x)\right)^T.
\end{aligned}
\end{equation}
{}From which we deduce that the flat coordinates $t^i$ are given by
\begin{equation}
\label{f5.9}
\theta_j(x,z)=\delta_{j,1}+
t^j(x)z+\sum_{k=2}^\infty\theta_j^{(k)}(x)z^k.
\end{equation}
These $\theta_j(x,z)$ are the deformed flat coordinates, see e.g. \cite{DuZ}.
So we are in the situation of the paper \cite{AKrV} and we can construct the
prepotential $F(t(x))$. Using the formula's (\ref{f5.4}) and (\ref{ff5.1})
one calculates that
\begin{equation}
\label{f5.10}
\begin{aligned}
\frac{\partial^2 \Theta(x,z)}{\partial x_1^{(i)}\partial x_1^{(j)}}&
=\Gamma_{ij}^i(x)\frac{\partial \Theta(x,z)}{\partial x_1^{(i)}}
\Gamma_{ji}^j(x)\frac{\partial \Theta(x,z)}{\partial x_1^{(j)}},\qquad i\ne j\\
\frac{\partial^2 \Theta(x,z)}{\partial x_1^{(i)2}}&
=\sum_{j=1}^n
\Gamma_{ii}^j(x)\frac{\partial \Theta(x,z)}{\partial x_1^{(j)}} +
z\frac{\partial^2 \Theta(x,z)}{\partial x_1^{(i)}\partial x_1^{(j)}},
\end{aligned}
\end{equation}
where the Christoffel symbols are given by (\ref{f09})
and hence that
\begin{equation}
\label{f5.11}
\frac{\partial^2 \Theta(x,z)}{\partial t^k \partial t^\ell}=
\sum_{m=1}^n c_{k\ell m}(x)z\frac{\partial \Theta(x,z)}{\partial t^m}
\end{equation}
Since
$\frac{\partial \Theta(x,z)}{\partial t^i}$ is a linear combination of
$\frac{\partial \Theta(x,z)}{\partial x^{(k)}_1}$'s,
\[
z^{-1} \frac{\partial \Theta(x,z)}{\partial t^k}\Theta(x,-z)^T
\]
is independent of $z$, which means that all coefficients, except the constant
coefficient, are zero. In particularly  the coefficient of $z^2$ gives:
\[
\theta_m^{(2)}(x)=-\frac{\partial \theta_1^{(3)}(x)}{t^m}+\sum_{i=1}^n
t^i\frac{\partial \theta_i^{(2)}(x)}{\partial t^m}.
\]
The coefficient  of $z^{2}$ of (\ref{f5.11}) leads to
\[
\frac{\partial^2 \theta_m^{(2)}(x)}{\partial t^k \partial t^\ell}=c_{k\ell
m}(x),
\]
hence $\frac{\partial F(x)}{\partial t^m}=\theta^{(2)}_m(x)$ and
we obtain the Theorem of \cite{AKrV} for the $GL(L)^{(2)}$-group orbit:
\begin{theorem}
The function $F(x)=F(t(x),x)$ defined by
\[
F(x)=-\frac{1}{2}\theta_1^{(3)}(x)+\frac{1}{2}\sum_{i=1}^n
t^i(x)\theta_i^{(2)}(x)
\]
satisfies equation (\ref{f01b}).
\end{theorem}

\end{document}